\global\def\draftcontrol{0}

   \def\versionno{ n2 on s4 }

\catcode`\@=11

\expandafter\ifx\csname draftcontrol\endcsname\relax\global\def\draftcontrol{0}
\fi

{\count255=\time\divide\count255 by 60
\xdef\hourmin{\number\count255}
\multiply\count255 by-60\advance\count255 by\time
\xdef\hourmin{\hourmin:\ifnum\count255<10 0\fi\the\count255}}
\def\draftdate{\number\month/\number\day/\number\year\ \ \ \hourmin }

\newcommand\makepapertitle{\par
  \begingroup
    \renewcommand\thefootnote{\@fnsymbol\c@footnote}%
    \def\@makefnmark{\rlap{\@textsuperscript{\normalfont\@thefnmark}}}%
    \long\def\@makefntext##1{\parindent 1em\noindent
            \hb@xt@1.8em{%
                \hss\@textsuperscript{\normalfont\@thefnmark}}##1}%
     \newpage
     \global\@topnum\z@   
     \@makepapertitle
     \thispagestyle{empty}\@thanks
  \endgroup
  \setcounter{footnote}{0}%
  \global\let\thanks\relax
  \global\let\makepapertitle\relax
  \global\let\@makepapertitle\relax
  \global\let\@thanks\@empty
  \global\let\@author\@empty
  \global\let\@date\@empty
  \global\let\@title\@empty
  \global\let\title\relax
  \global\let\author\relax
  \global\let\date\relax
  \global\let\and\relax
  \def\version{\let\version\@version\@gobble}
}
\def\@makepapertitle{%
  \newpage
   \ifnum\draftcontrol=1 {}
   \version\versionno
   \vskip 3em%
   \else
   \hfill\hbox to 3cm {\parbox{4cm}{\@pubnum}\hss}%
   \vskip 3em%
   \fi
   \begin{center}%
   \let \footnote \thanks
     {\LARGE {\@title}}%
     \vskip 1.5em%
     {\normalsize
       \lineskip .5em%
       \begin{tabular}[t]{c}%
         \@author
       \end{tabular}\par}%
     \vskip 1.5em%
     {\@bstract}%
     \end{center}%
     \vskip 1.5em
     \@date%
   \par
}

\gdef\@pubnum{}
\def\pubnum#1{%
  \gdef\@pubnum{#1}}

\gdef\@bstract{}
\def\Abstract#1{%
  \gdef\@bstract{%
   \parbox{\textwidth-0pc}{%
   \centerline{\bf Abstract}\penalty1000%
\kern.2cm%
\noindent
\renewcommand\baselinestretch{1.0}%
{#1}}}
}

\def\ps@paper{\let\@mkboth\@gobbletwo%
     \ifnum\draftcontrol=1
    \def\@oddfoot{\hbox to \textwidth{\tiny \versionno \hfil\tiny\draftdate}%
    \hskip -\textwidth \hbox to \textwidth{\hfil\rm\thepage\hfil}}%
     \else\def\@oddfoot{\hbox to \textwidth{\hfil\rm\thepage\hfil}}
     \fi
     \let\@evenfoot\@oddfoot
}

\def\body{\clearpage
          \pagestyle{paper}
    }

\def\@version#1{\ifnum\draftcontrol=1
\typeout{}\typeout{#1}\typeout{}
\vskip3mm\centerline{\hbox{\fbox{\normalsize{\tt DRAFT -- #1 -- }
                   {\draftdate}}}}\vskip3mm
\fi}
\let\version\@version
\long\def\eqlabel#1{\ifnum\draftcontrol=1
                    \tag@false  
                    \tag*{(\theequation) \hbox to -0.2cm{\hspace{0cm}\small{#1}\hss}}
                    \refstepcounter{equation}
                    \edef\@currentlabel{\theequation}
                    \ltx@label{#1}          
                    \else
                    \label{#1}
                    \fi
                    }
\let\st@bibitem\@bibitem
\let\st@lbibitem\@lbibitem
\ifnum\draftcontrol=1
  \def\@bibitem#1{%
    \st@bibitem{#1}\a@@label{#1}\ignorespaces}
  \def\@lbibitem[#1]#2{%
    \st@lbibitem[#1]{#2}\a@@label{#2}\ignorespaces}
  \def\a@@label#1{%
    \gdef\a@lab{\smash{\normalfont\small#1}}
    \ifvmode
      \if@inlabel
        \global\setbox\@labels\hbox{%
          \llap{\a@lab\let\a@lab\relax
                \kern\@totalleftmargin\kern\marginparsep}%
          \box\@labels}%
      \fi
    \fi}
\fi

\documentclass[12pt,letterpaper]{article}

\usepackage{amsmath,amssymb,array,calc,epsfig,rotating,bm}
\usepackage[sort]{cite}
\usepackage{graphicx}
\usepackage{psfrag,verbatim,MnSymbol}

\ifnum\draftcontrol=1
\fi

\renewcommand\baselinestretch{1.25}
\renewcommand\section{\@startsection {section}{1}{\z@}%
                                   {-3.5ex \@plus -1ex \@minus -.2ex}%
                                   {2.3ex \@plus.2ex}%
                                   {\normalfont\large\bfseries}}
\renewcommand\subsection{\@startsection{subsection}{2}{\z@}%
                                   {-3.25ex\@plus -1ex \@minus -.2ex}%
                                   {1.5ex \@plus .2ex}%
                                   {\normalfont\normalsize\bfseries}}
\renewcommand\subsubsection{\@startsection{subsubsection}{3}{\z@}%
                                   {-3.25ex\@plus -1ex \@minus -.2ex}%
                                   {1.5ex \@plus .2ex}%
                                   {\normalfont\normalsize\it}}
\renewcommand\paragraph{\@startsection{paragraph}{4}{\z@}%
                                   {-3.25ex\@plus -1ex \@minus -.2ex}%
                                   {1.5ex \@plus .2ex}%
                                   {\normalfont\normalsize\bf}}

\numberwithin{equation}{section}

\def\revise#1       {\raisebox{-0em}{\rule{3pt}{1em}}%
                     \marginpar{\raisebox{.5em}{\vrule width3pt\
                     \vrule width0pt height 0pt depth0.5em
                     \hbox to 0cm{\hspace{0cm}{%
                     \parbox[t]{4em}{\raggedright\footnotesize{#1}}}\hss}}}}

\newcommand\nxt[1]  {\\\fnxt#1}
\newcommand{\ie}{{\it i.e.,}\ }

\def\cala         {{\cal A}}

\def\calc         {{\cal C}}

\def\calf         {{\cal F}}

\def\cali         {{\cal I}}
\def\calj         {{\cal J}}

\def\call         {{\cal L}}
\def\calm         {{\cal M}}
\def\caln         {{\cal N}}
\def\calo         {{\cal O}}
\def\calp         {{\cal P}}

\def\cals         {{\cal S}}

\def\calz         {{\cal Z}}

\def\complex      {{\mathbb C}}

\def\del          {\partial}

\def\Im           {{\rm Im\hskip0.1em}}

\def\sqr#1#2{{\vcenter{\vbox{\hrule height.#2pt
 \hbox{\vrule width.#2pt height#1pt \kern#1pt
 \vrule width.#2pt}\hrule height.#2pt}}}}

\newcommand{\fft}[2]{{\frac{#1}{#2}}}
\newcommand{\ft}[2]{{\textstyle{\frac{#1}{#2}}}}
\def\jsquare{\mathop{\mathchoice{\sqr{8}{32}}{\sqr{8}{32}}
{\sqr{6.3}{9}}{\sqr{4.5}{9}}}}

\def\a{\alpha}

\def\r{\rho}
\def\dd{\delta}
\def\e{\epsilon}
\def\c{\chi}

\def\aa1{\phi}
\def\cc1{\psi}

\def\arctanh{{\rm arctanh}}

\def\l{\lambda}

\def\ha{\hat{a}}

\def\t{\tau}
\def\tl{\tilde{\Lambda}}

\def\arcsinh{{\rm arcsinh}}

\catcode`\@=12

\begin{document}


\title{\bf   Localization and holography in  $\caln=2$ gauge theories}
\pubnum{UWO-TH-13/7}

\date\today

\author{
Alex Buchel\\[0.4cm]
\it Perimeter Institute for Theoretical Physics\\
\it Waterloo, Ontario N2J 2W9, Canada\\[0.2cm]
\it Department of Applied Mathematics, University of Western Ontario\\
\it London, Ontario N6A 5B7, Canada\\[0.2cm]
}

\Abstract{
We compare exact results from Pestun's localization
\cite{Pestun:2007rz} of $SU(N)$ $\caln=2^*$ gauge theory on $S^4$ 
with available holographic models. While localization can explain the
Coulomb branch vacuum of the holographic Pilch-Warner flow \cite{pw},
it disagrees with the holographic Gauntlett et.al
\cite{Gauntlett:2001ps} vacuum of $\caln=2$ super Yang-Mills theory.
We further compute the free energy of the Pilch-Warner flow on $S^4$
and show that it disagrees with the localization result both for a
finite $S^4$ radius, and in the $S^4$ decompactification limit. Thus,
neither model represents holographic dual of supersymmetric $S^4$
localization of \cite{Pestun:2007rz}.
}

\makepapertitle

\body

\version\versionno
\tableofcontents

\section{Introduction and motivation}
Original work of Maldacena \cite{m1} established a duality between $\caln=4$ 
$SU(N)$ supersymmetric Yang-Mills (SYM) theory and 
type IIb string theory. A lot 
of subsequent work was devoted to generalizing the holographic correspondence 
to non-conformal theories, and theories with reduced supersymmetry. 
Extensions of gauge/gravity correspondence to gauge 
theories with eight supercharges (the $\caln=2$ supersymmetric models)
play a special role, as they allow for a direct check of the 
correspondence with the field-theoretic Seiberg-Witten solution 
\cite{Seiberg:1994rs,Argyres:1994xh}. The two notable examples 
of $\caln=2$ dualities are the Pilch-Warner
\cite{pw} (PW) and the Gauntlett et.al 
\cite{Gauntlett:2001ps} (GKMW) holographic renormalization 
group (RG) flows. In the former, one considers a planar limit of 
$\caln=4$ $SU(N)$ SYM at large 't Hooft coupling, deformed by 
$\caln=2$ hypermultiplet mass term (the so called $\caln=2^*$ 
gauge theory); in the latter, one starts with the $S^2$-compactified 
Little String Theory in the ultraviolet, and flows in the infrared to   
large-$N$ $\caln=2$ $SU(N)$ SYM. $\caln=2$ gauge theories have 
quantum Coulomb branch vacua $\calm_{\calc}$, parameterized by the expectation 
values of the complex scalar $\Phi$ in the $\caln=2$ vector multiplet, 
taking values in the Cartan subalgebra of the gauge group. For the 
$SU(N)$ gauge group,
\begin{equation}
\Phi={\rm diag}(a_1,a_2,\cdots,a_N)\,,\qquad \sum_i a_i=0\,,
\eqlabel{cvev}
\end{equation}
resulting in complex dimension of the moduli space
\begin{equation}
{\rm dim}_\complex\ \calm_\calc\ =\ N-1\,. 
\eqlabel{dim}
\end{equation}
In the large-$N$ limit, and for strong 't Hooft coupling, the holographic 
duality reduces to the correspondence between the gauge theory and 
type IIb supergravity. Since supergravities have finite number of light modes, 
one should not expect to see the full moduli space of vacua in $\caln=2$ 
examples of gauge/gravity correspondence. This is indeed what is happening:
the PW flow localizes on a semi-circle distribution of \eqref{cvev} with a
linear number density  \cite{bpp},
\begin{equation}
\begin{split}
&\Im(a_i)=0\,,\qquad a_i\in [-a_0,a_0]\,,\qquad a_0^2=\frac{m^2 g_{YM}^2 N}{4\pi^2}\,,\\
&\r(a)=\frac{8\pi}{m^2 g_{YM}^2}\ \sqrt{a_0^2-a^2}\,,\qquad \int_{-a_0}^{a_0}da\ 
\r(a)=N\,,
\end{split}
\eqlabel{pwdistr}
\end{equation}
where $m$ is the hypermultiplet mass; the GKMW flow localizes on a circular 
distribution of eigenvalues, centered about the origin 
with radius $u_0\propto \sqrt{N}$ \cite{Gauntlett:2001ps},  
\begin{equation}
\begin{split}
&\r(a)=\frac{N}{2\pi u_0}\ \dd(|a|-u_0)\,,\qquad \int\int_{\complex} d^2a\ 
\r(a)=N\,.
\end{split}
\eqlabel{gkmwdistr}
\end{equation}
An outstanding question is the mechanism of the holographic localization on the moduli 
space of $\caln=2$ vacua in the large-$N$ limit.

A possible mechanism of holographic localization was proposed in 
\cite{Buchel:2013id}\footnote{See also \cite{Russo:2012ay,Russo:2013qaa}.}. 
In \cite{Pestun:2007rz} Pestun pointed out that the partition function of $\caln=2^*$ gauge theory on 
$S^4$, and some supersymmetric observables, can be computed exactly from the corresponding matrix model.
This $S^4$ compactification does not twist the supersymmetry --- around the trivial background the 
compactified theory does not have zero modes. In the large-$N$ limit, the gauge theory partition function   
naturally localizes on the saddle-point of the corresponding matrix model, thus providing a selection 
mechanism for the Coulomb branch vacuum in the $S^4$ decompactification limit. Such argument
indeed explains the PW vacuum \eqref{pwdistr}, as well as correctly reproduces
the holographic computation of the supersymmetric Wilson loop in PW geometry \cite{pw}.
So, does it mean that $\caln=2$ PW consistent truncation of type IIb supergravity contain 
large-$N$ holographic dual to Pestun's supersymmetric compactification of $\caln=2^*$ gauge theory 
on $S^4$? 

To answer this question we compute the free energy of the compactified Euclidean 
PW flow on the four-sphere of radius $R$. Ambiguities of the holographic renormalization 
imply that for  $m R\ll 1 $ only $\calo(m^6 R^6)$ (and subleading terms) 
are renormalization-scheme independent. We show that at order $\calo(m^6 R^6)$ there is a disagreement 
between the holographic and the matrix model free energies. Such disagreement indicates that PW truncation 
is inconsistent with the $\caln=2^*$ gauge theory $S^4$ supersymmetries of \cite{Pestun:2007rz}.  
Still, motivated by the success of \cite{Buchel:2013id}, the possibility remains that the holographic and
the matrix model free energies agree in the $S^4$ decompactification limit. Unfortunately, 
we study the free energy of $S^4$-compactified PW flow in the limit $m R\gg 1$ and find 
that it is in conflict with the matrix model result. Thus, Pestun's localization 
can not explain holographic localization of $\caln=2$ supergravity flows. The last point is further 
stressed by pointing out that large-$N$ localization of $\caln=2$ $SU(N)$ SYM \cite{Russo:2012ay} 
is different from the holographic vacuum localization of GKMW supergravity flow. 

In section \ref{section2} we collect relevant localization results for $\caln=2^*$ gauge theory 
\cite{Buchel:2013id}, and for $\caln=2$ SYM \cite{Russo:2012ay}. While the corresponding 
matrix model to $\caln=2^*$ gauge theory  localizes on holographic PW vacuum \eqref{pwdistr}, 
it fails to localize on GKMW vacuum \eqref{gkmwdistr} in the case of $\caln=2$ SYM. 
In section \ref{section3} we compute the free energy of $S^4$-compactified holographic 
PW flow. We compare the results with the matrix model computation. We conclude in section
\ref{section4}.     
 
{\bf Note added:} After the paper was published, I was informed by
Francesco Bigazzi about his relevant work
\cite{Bigazzi:2001aj} (see also \cite{Bigazzi:2003ui} 
). The authors of \cite{Bigazzi:2001aj} consider a more general ansatz
than GKMW for wrapped D5-branes, resulting in a {\it different}
$\caln=2$ Coulomb branch vacuum of large-$N$ $SU(N)$ SYM.  Unlike the
GKMW vacuum, this vacuum does agree with the localization vacuum
of \cite{Russo:2012ay}. It would be interesting to compare the
expectation values of circular Wilson loops and the free energy in
supergravity background of \cite{Bigazzi:2001aj} with the computations
in \cite{Russo:2012ay}.

\section{Localization of $\caln=2$ gauge theories on $S^4$ }\label{section2}

According to \cite{Pestun:2007rz}, the partition function of $\caln=2$ gauge theories on $S^4$ of 
radius $R$
reduces to $(N-1)$ dimensional integral over $\{\ha_i\equiv a_i R\}$ (see \eqref{cvev}) 
of an effective matrix model:
\nxt for $\caln=2^*$ $SU(N)$ gauge theory, $Z_{\caln=2^*}$,
\begin{equation}
Z_{\caln=2^*}=\int d^{N-1}\ha \prod_{i<j}\ \frac{(\ha_i-\ha_j)^2 H^2(\ha_i-\ha_j)}
{H(\ha_i-\ha_j-m R) H(\ha_i-\ha_j+m R)}\ e^{-\frac{8\pi^2 N}{\l}\sum_{j}\ha_j^2}
\ |\calz_{{\rm inst}}|^2\,,
\eqlabel{zn2star}
\end{equation}
with  $\l\equiv g_{YM}^2 N$;
\nxt for $\caln=2$ $SU(N)$ SYM, $Z_{SYM}$,
\begin{equation}
Z_{SYM}=\int d^{N-1}\ha \prod_{i<j}\ \left[{(\ha_i-\ha_j)^2 H^2(\ha_i-\ha_j)}
\right]\ e^{-\frac{8\pi^2 N}{\l}\sum_{j}\ha_j^2}
\ |\calz_{{\rm inst}}|^2\,,
\eqlabel{zsym}
\end{equation}
with the running 't Hooft coupling $\l$ evaluated at the cut-off set by the 
$S^4$ radius $R$, \cite{Russo:2012ay}:
\begin{equation}
\frac{4\pi^2}{\l}=-\ln(\Lambda R)\,.
\eqlabel{lsym}
\end{equation}
In \eqref{lsym} $\Lambda$ is the strong coupling scale of the SYM.
The function $H(x)$ is expressed as an infinite product over the spherical harmonics,
\begin{equation}
H(x)\equiv \prod_{n=1}^\infty\left(1+\frac{x^2}{n^2}\right)^n\ e^{-\frac{x^2}{n}}\,.
\eqlabel{defh}
\end{equation}

Matrix integrals \eqref{zn2star} and \eqref{lsym} dramatically simplify 
in the large-$N$ limit. First,  as the instantons are suppressed in the planar limit, 
we can set $|\calz_{inst}|=1$. Second, the saddle point approximation becomes exact \cite{Russo:2012kj}. 

In what follows, it is convenient to introduce 
\begin{equation}
K(x)\equiv -\left(\ln H(x)\right)' \,.
\eqlabel{defk}
\end{equation}
We further set $R=1$ (and drop the caret)--- along with $m$ (for $\caln=2^*$ gauge theory) or $\Lambda$
(for the SYM) the $S^4$ radius is the only other dimensionful scale; thus the $R$-dependence can always be 
restored from dimensional analysis.

\subsection{$\caln=2^*$ gauge theory}

The saddle-point equations derived from \eqref{zn2star} take form  \cite{Russo:2012kj,Buchel:2013id}
\begin{equation}
\frac 1N \sum_{k\ne j} \left(\frac{1}{a_j-a_k}-K(a_j-a_k)+\frac 12 
K(a_j-a_k+m)+\frac 12 K(a_j-a_k-m) \right)=\frac{8\pi^2}{\l}\ a_j\,.
\eqlabel{n2eq1}
\end{equation}
Assuming\footnote{This is justified \it{a posteriori}.} $\Im(a_i)=0$, $a_i\in[-\mu,\mu]$, and introducing 
a linear eigenvalue density 
\begin{equation}
\r(x)=\frac 1N\ \sum_i \dd(x-a_i)\,,\qquad \int_{-\mu}^\mu dx\ \r(x)=1\,,
\eqlabel{defrho}
\end{equation}
we find 
\begin{equation}
\strokedint_{-\mu}^\mu dy\, 
\rho(y) \left(\frac{1}{x-y} -K(x-y)+\frac12\,K(x-y+m)+\frac{1}{2}\,K(x-y-m)\right)= \frac{8\pi^2}{\lambda}\, x\,.
\eqlabel{n2eq2}
\end{equation}
In the limit\footnote{As emphasized in \cite{Russo:2013qaa} this limit 
has an irregular fuzzy fine structure.} $\l\to \infty$ the solution is given by \cite{Buchel:2013id}
\begin{equation}
\r(x)=\frac{2}{\pi\mu^2}\sqrt{\mu^2-x^2}\,,\qquad \mu=\frac{\sqrt{\l(\frac{1}{R^2}+m^2)}}{2\pi}\,,
\eqlabel{n2rho}
\end{equation}
where we restored the $R$-dependence. In the $S^4$ decompactification limit, \ie  $m R\to \infty$, 
the distribution \eqref{n2rho} reproduces  the PW vacuum  \eqref{pwdistr}. 

The free energy $\calf_{\caln=2^*}^{loc}$, 
\begin{equation}
\calf_{\caln=2^*}^{loc}=-\ln Z_{\caln=2^*}\,,
\eqlabel{fn2eq1}
\end{equation}
can be computed by first differentiating it with respect to $m$:
\begin{equation}
\begin{split}
\frac{\del}{\del m}\ \calf_{\caln=2^*}^{loc}&=\bigg\langle \frac 12 \sum_{i,j} \biggl
(K(a_i-a_j-m)-K(a_i-a_j-m)\biggr)\bigg\rangle\\
&=\frac{N^2}{2}\ \int\int dxdy\ \r(x)\r(y)\ \biggl(K(x-y-m)-K(x-y+m)\biggr)\,.
\end{split}
\eqlabel{fn2eq2}
\end{equation}
The leading contribution in the limit $\l\to\infty$ then becomes \cite{Buchel:2013id}
\begin{equation}
\frac{\del}{\del m}\ \calf_{\caln=2^*}^{loc}=-N^2 m\ \ln\ \frac{\l(1+m^2)e^{2\gamma+\frac 32}}{16\pi^2}\,,
\eqlabel{fn2eq3}
\end{equation}
where $\gamma=-\psi(1)$ is the Euler's constant. Ensuring that the free energy 
agrees with that of the $\caln=4$ SYM  in the limit $m\to 0$ \cite{Russo:2012ay} we find, 
\begin{equation}
\calf_{\caln=2^*}^{loc}=-\frac{N^2}{2} (1+m^2 R^2)\ \ln\ \frac{\l(1+m^2R^2)e^{2\gamma+\frac 32}}{16\pi^2}\,.
\eqlabel{fn2eq4}
\end{equation}
From \eqref{fn2eq4} it is easy to extract the small- and large-$R$ limits:
\begin{equation}
\calf_{\caln=2^*}^{loc}={N^2} \ \times\ 
\begin{cases}
\calc+\left(-\frac 12+\calc\right)m^2 R^2-\frac 14 m^4R^4+\frac{1}{12} m^6 R^6+\calo(m^8R^8)\,,\qquad (mR\ll 1)\,, \\
- m^2 R^2 \ln (mR)\,,\qquad (m R\gg 1)\,,
\end{cases}
\eqlabel{fneq5}
\end{equation}
where we denoted 
\begin{equation}
\calc=-\frac 12 \ln\ \frac{\l e^{2\gamma+\frac 32}}{16\pi^2}\,.
\eqlabel{fneq6}
\end{equation}
In section \ref{section3} we compute the free energy of $S^4$-compactified PW flow. 
We find that the holographic free energy disagrees with \eqref{fn2eq4}, \eqref{fneq5}. 

\subsection{$\caln=2$ SYM}

The saddle-point equations derived from \eqref{zsym} take form  \cite{Russo:2012ay}
\begin{equation}
\frac 1N \sum_{k\ne j} \left(\frac{1}{a_j-a_k}-K(a_j-a_k)\right)=\frac{8\pi^2}{\l}\ a_j=-2 a_j\ \ln\Lambda\,.
\eqlabel{nsymeq1}
\end{equation}
Assuming\footnote{Again, this is justified \it{a posteriori}.} $\Im(a_i)=0$, $a_i\in(-\mu,\mu)$, and introducing 
a linear eigenvalue density as in \eqref{defrho}, we find 
\begin{equation}
\strokedint_{-\mu}^\mu dy\, 
\rho(y) \left(\frac{1}{x-y} -K(x-y)\right)= -2x\ \ln\Lambda\,.
\eqlabel{nsymeq2}
\end{equation}
Analytic solutions to \eqref{nsymeq2} are available in the limiting cases $\Lambda\ll 1$
and $\Lambda\gg 1$ \cite{Russo:2012ay}. As we will be interested in the $S^4$ decompactification limit,
we will focus on the latter case:
\begin{equation}
\r(x)=\frac{1}{\pi \sqrt{\mu^2 -x^2}}\,,\qquad \mu=2 e^{-1-\gamma}\ \Lambda R\,.
\eqlabel{distsym}
\end{equation}  
Solution \eqref{distsym} has to be taken with a grain of salt:
\nxt first, the boundary conditions for the eigenvalue density $\rho(x)$ are not satisfied, \ie 
\begin{equation}
\lim_{x\to \pm \mu}\r(x)=0\,;
\eqlabel{bcrho}
\end{equation}
\nxt second, the $S^4$ decompactification limit is problematic, as 
in this case $\lambda\to 0_- $ (see \eqref{lsym}).
We return to these points later.

The free energy $\calf_{SYM}^{loc}$, 
\begin{equation}
\calf_{SYM}^{loc}=-\ln Z_{SYM}\,,
\eqlabel{fsymeq1}
\end{equation}
can be computed by first differentiating it with respect to $\Lambda$:
\begin{equation}
\begin{split}
\frac{\del}{\del \ln \Lambda}\ \calf_{SYM}^{loc} &=-2 N\ \bigg\langle \sum_j a_j^2\bigg\rangle\\
&=-2 N^2\ \int dx\ \r(x)\ x^2\\
&=-N^2 \mu^2=-4 e^{-2 -2\gamma} N^2 \Lambda^2 R^2\,,\qquad R \Lambda\gg 1\,,
\end{split}
\eqlabel{fsymeq2}
\end{equation}
leading to 
\begin{equation}
\calf_{SYM}^{loc}=- 2 e^{-2 -2\gamma} N^2 \Lambda^2 R^2\,,\qquad R \Lambda\gg 1\,.
\eqlabel{fsymeq3}
\end{equation}

We would like to compare matrix model results of \cite{Russo:2012ay} reviewed above with the
holographic computation of GKMW \cite{Gauntlett:2001ps}. In the latter holographic RG flow 
one starts with large number of NS5 branes wrapping a 2-cycle in the UV, and flows in the 
IR to $SU(N)$ SYM. The strong coupling scale of the SYM is set by the $S^2-$compactification scale
of Little String Theory on the world-volume of the five-branes. Clearly, the field theory dual to GKMW 
flow at high-energy can not be a SYM. Thus, there is no reason to expect that the small $S^4$ radius localization 
results would agree with results from  the small-$R$ $S^4$-compactified GKMW flow. One would expect, however,
that the moduli space properties of GKMW would agree with the matrix model computations in the 
limit $\Lambda R\to \infty$.  Unfortunately, this is not the case: compare \eqref{gkmwdistr} and \eqref{distsym}.
We can identify the disagreement a bit more precisely. 
Following \cite{Gauntlett:2001ps}, the probe $U(1)$ gauge coupling $\tau$ on the GKMW moduli space
parameterized by $u$ takes the form
\begin{equation}
\t(u)=i\ \frac{2N}{\pi}\ \ln \frac{u}{\tl}\,,
\eqlabel{tausym}
\end{equation}       
where $\tl$ is the IR cutoff, related to the strong coupling scale of the SYM.
The $\caln=2$ supersymmetry relates the coupling to the metric on the moduli space 
as\footnote{We use $R$ simply as a dimensionful parameter to facilitate the comparison with the 
matrix model result \eqref{distsym}. } 
\begin{equation}
\begin{split}
\t(u)=&i\ \frac{N}{\pi}\ \int dx \r(x)\ \ln\frac{(u R-x)^2}{{\mu}^2} \\
=& i\ \frac{2 N}{\pi}\ \ln \frac{\sqrt{u^2 R^2 -\mu^2}+u R}{2\mu}\,,\qquad u R> \mu\,,
\end{split}
\eqlabel{rel1}
\end{equation}
where we used the saddle-point vacuum of the matrix model 
\eqref{distsym}. 
Notice that \eqref{tausym} and \eqref{rel1} agree in the limit $u\gg \Lambda$, provided we relate 
\begin{equation}
\tl=2 e^{-1-\gamma} \Lambda\,,
\eqlabel{idtl}
\end{equation} 
but disagree in general. This discrepancy might be attributed to the fact that the RG equation 
in the matrix model \eqref{lsym} makes sense only up to $R\sim \frac 1\Lambda$, which in 
turn implies that the moduli space coupling \eqref{rel1} can be accurate only 
when evaluated at $|u|\gg \frac 1R\sim \Lambda$. 

Having established that the Coulomb branch vacua in the 
saddle-point matrix model for $SU(N)$ SYM and the GKMW flow differ,
there is no point to proceed with the detailed comparison of the corresponding 
free energies.

\section{$\caln=2^*$ free energy on $S^4$ from holography}\label{section3}
In this section we compute the free energy $\calf$ of  
$S^4$-compactified Pilch-Warner holographic RG flow. 
We compare the latter with
the matrix model result $\calf_{\caln=2^*}^{loc}$ \eqref{fn2eq4} --- we stress that here, 
unlike the relation between the $\caln=2$ SYM and the GKMW holographic RG 
flow, it makes sense to compare the free energies for generic values of 
$mR$.

\subsection{Effective action and equations of motion}\label{solution}
The supergravity background dual to $S^4$ compactification 
of  $\caln=2^*$ gauge theory \cite{b1} is a deformation
of the original $AdS_5\times S^5$ geometry\footnote{With $S^4$ slicing of $AdS_5$ 
--- see \cite{b2}. } induced by a pair of
scalars $\alpha$ and $\chi$ of the five-dimensional gauge
supergravity. (At zero temperature, such a deformation was
constructed by Pilch and Warner (PW) 
\cite{pw}\footnote{See \cite{bpp,cj} for the gauge theory 
interpretation of the PW geometry.}.) According to the
general scenario of a holographic RG flow, the asymptotic
boundary behavior of the supergravity scalars is related to the
bosonic and fermionic mass parameters of the relevant operators
inducing the RG flow in the boundary gauge theory. Based on such a
relation, and the fact that $\alpha$ and $\chi$ have conformal dimensions
two and one, respectively, we call the supergravity 
scalar $\a$ a {\it bosonic} deformation, and the supergravity scalar $\chi$  
a {\it fermionic} deformation of the D3-brane geometry.

The action of the effective five-dimensional gauged supergravity including the
scalars $\alpha$ and $\chi$ is given by
\begin{equation}
\begin{split}
S=&\,
\int_{\calm_5} d\xi^5 \sqrt{-g}\ \call_5\\
=&\frac{1}{4\pi G_5}\,
\int_{\calm_5} d\xi^5 \sqrt{-g}\left[\ft14 R-3 (\del\a)^2-(\del\chi)^2-
\calp\right]\,,
\end{split}
\eqlabel{action5}
\end{equation}
where the potential%
\footnote{We set the five-dimensional gauged
supergravity coupling to one. This corresponds to setting the
radius $L$ of the five-dimensional sphere in the undeformed metric
to $2$.}
\begin{equation}
\calp=\frac{1}{16}\left[\frac 13 \left(\frac{\del W}{\del
\a}\right)^2+ \left(\frac{\del W}{\del \chi}\right)^2\right]-\frac
13 W^2\,
 \eqlabel{pp}
\end{equation}
is a function of $\alpha$ and $\chi$, and is determined by the
superpotential
\begin{equation}
W=- e^{-2\alpha} - \frac{1}{2} e^{4\alpha} \cosh(2\chi)\,.
\eqlabel{supp}
\end{equation}
In our conventions, the five-dimensional Newton's constant is
\begin{equation}
G_5\equiv \frac{G_{10}}{2^5\ {\rm vol}_{S^5}}=\frac{4\pi}{N^2}\,.
\eqlabel{g5}
\end{equation}
The action \eqref{action5} yields the Einstein equations
\begin{equation}
R_{\mu\nu}=12 \del_\mu \a \del_\nu\a+ 4 \del_\mu\chi\del_\nu\chi
+\frac43 g_{\mu\nu} \calp\,,
\eqlabel{ee}
\end{equation}
as well as the equations for the scalars
\begin{equation}
\jsquare\alpha=\fft16\fft{\del\calp}{\del\alpha}\,,\qquad
\jsquare\chi=\fft12\fft{\del\calp}{\del\chi}\,.
\eqlabel{scalar}
\end{equation}

To construct the  $S^4$  compactification of the Pilch-Warner
flow, we choose an ansatz for the metric respecting 
$SO(5)$ rotational invariance
\begin{equation}
ds_5^2=c_1^2(r)\ \left(dS^4\right)^2+ \ dr^2\,,
\eqlabel{ab}
\end{equation}
where $\left(dS^4\right)^2$ is a round metric on $S^4$ of unit radius.
With this ansatz, the equations of motion for the background become
\begin{equation}
\begin{split}
&\a''+\a'\ \left(\ln{c_1^4}\right)'
-\frac 16 \frac{\del\calp}{\del\a}=0\,,\\
&\c''+\c'\ \left(\ln{c_1^4}\right)'-\frac 12 \frac{\del\calp}{\del\c}=0\,,\\
&c_1''+c_1'\ \left(\ln{c_1^3}\right)'-\frac 3c_1+\frac 43c_1  \calp=0\,,
\end{split}
\eqlabel{beom}
\end{equation}
where the prime denotes a derivative with respect to the radial
coordinate $r$.
In addition, there is a first-order constraint
\begin{equation}
\left(\a'\right)^2+\frac 13 \left(\c'\right)^2 -\frac 13
\calp- \left((\ln c_1)'\right)^2+\frac{1}{c_1^2} =0\,. 
\eqlabel{backconst}
\end{equation}
It was shown in \cite{b1} that any solution to \eqref{beom} and
\eqref{backconst} can be lifted to a 
full ten-dimensional solution of type IIb supergravity. This includes 
the metric, the three- and five-form fluxes, the dilaton and the axion.
In particular, the ten-dimensional 
Einstein frame metric is given by Eq.~(32) in \cite{b1}.

For $S^4$-compactified flow, we find it convenient to introduce a
new radial coordinate $x$ as follows :
\begin{equation}
x(r) = c_1(r)\,, \qquad x\in [0,+\infty)\,.
\eqlabel{radgauge}
\end{equation}
With this new coordinate, 
the background equations of motion \eqref{beom} become
\begin{equation}
\begin{split}
&0=\r''-4 x (-144 \r^4 c^4+\r^{12} x^2 c^8-16 \r^6 x^2 c^6-2 \r^{12} x^2 c^4+\r^{12} x^2\\
&-16 x^2 c^4-16 \r^6 x^2 c^2) (\r')^3/(\r^2 (-16 \r^6 x^2 c^2-192 \r^4 c^4-16 x^2 c^4
+\r^{12} x^2 c^8-16 \r^6 x^2 c^6\\
&-2 \r^{12} x^2 c^4+\r^{12} x^2))-(-32 \r^6 x^2 c^2+16 x^2 c^4-192 \r^4 c^4+5 \r^{12} x^2 c^8-32 \r^6 x^2 c^6\\
&-10 \r^{12} x^2 c^4+5 \r^{12} x^2) (\r')^2/(\r (-16 \r^6 x^2 c^2-192 \r^4 c^4-16 x^2 c^4
+\r^{12} x^2 c^8-16 \r^6 x^2 c^6\\
&-2 \r^{12} x^2 c^4+\r^{12} x^2))
+(1/3) (-240 \r^6 x^2 c^4-240 x^2 c^6+15 \r^{12} x^2 c^{10}-240 \r^6 x^2 c^8
\\
&-30 \r^{12} x^2 c^6+15 \r^{12} x^2 c^2-4 (c')^2 \r^{12} x^4-2304 \r^4 c^6+64 (c')^2 \r^6 x^4 c^2+576 \r^4 (c')^2 c^4 x^2\\
&+64 (c')^2 c^4 x^4+8 (c')^2 c^4 \r^{12} x^4+64 (c')^2 c^6 \r^6 x^4-4 (c')^2 c^8 \r^{12} x^4) \r'/(x c^2 (-16 \r^6 x^2 c^2\\
&-192 \r^4 c^4-16 x^2 c^4+\r^{12} x^2 c^8-16 \r^6 x^2 c^6-2 \r^{12} x^2 c^4+\r^{12} x^2))+(4/3) \r (-x^2 \r^{12} (c')^2\\
&-6 \r^{12} c^6+3 \r^{12} c^2+3 \r^{12} c^{10}+24 c^6-12 \r^6 c^4-12 \r^6 c^8-8 x^2 (c')^2 c^4-x^2 \r^{12} (c')^2 c^8\\
&+2 x^2 \r^{12} (c')^2 c^4+4 x^2 \r^6 (c')^2 c^6+4 x^2 \r^6 (c')^2 c^2)/(c^2 (-16 \r^6 x^2 c^2-192 \r^4 c^4-16 x^2 c^4\\
&+\r^{12} x^2 c^8-16 \r^6 x^2 c^6-2 \r^{12} x^2 c^4+\r^{12} x^2))\,,\\
\end{split}
\eqlabel{beomx1}
\end{equation}
\begin{equation}
\begin{split}
&0=c''-(4/3) x (-144 \r^4 c^4+\r^{12} x^2 c^8-16 \r^6 x^2 c^6-2 \r^{12} x^2 c^4+\r^{12} x^2
-16 x^2 c^4\\
&-16 \r^6 x^2 c^2) (c')^3/(c^2 (-16 \r^6 x^2 c^2-192 \r^4 c^4-16 x^2 c^4+\r^{12} x^2 c^8
-16 \r^6 x^2 c^6-2 \r^{12} x^2 c^4\\
&+\r^{12} x^2))-(-32 \r^6 x^2 c^6+3 \r^{12} x^2 c^8-192 \r^4 c^4-16 x^2 c^4-2 \r^{12} x^2 c^4-\r^{12} x^2) (c')^2/(c (\\
&-16 \r^6 x^2 c^2-192 \r^4 c^4-16 x^2 c^4+\r^{12} x^2 c^8-16 \r^6 x^2 c^6-2 \r^{12} x^2 c^4+\r^{12} x^2))-(80 c^4 \r^2 x^2\\
&-64 c^4 (\r')^2 x^4-5 c^8 \r^{14} x^2+10 c^4 \r^{14} x^2+80 c^6 \r^8 x^2+80 c^2 \r^8 x^2-5 \r^{14} x^2+768 \r^6 c^4\\
&-576 c^4 \r^4 (\r')^2 x^2-64 c^6 (\r')^2 x^4 \r^6-8 c^4 (\r')^2 x^4 \r^{12}
+4 c^8 (\r')^2 x^4 \r^{12}-64 c^2 (\r')^2 x^4 \r^6\\
&+4 (\r')^2 x^4 \r^{12}) c'/(\r^2 x (-16 \r^6 x^2 c^2-192 \r^4 c^4
-16 x^2 c^4+\r^{12} x^2 c^8-16 \r^6 x^2 c^6-2 \r^{12} x^2 c^4\\
&+\r^{12} x^2))+6 \r^4 c (8 \r^2 c^2-8 \r^2 c^6+\r^8 c^8
-\r^8+x^2 \r^6 (\r')^2+8 x^2 c^6 (\r')^2-8 x^2 c^2 (\r')^2\\
&-x^2 \r^6 c^8 (\r')^2)/(-16 \r^6 x^2 c^2-192 \r^4 c^4-16 x^2 c^4+\r^{12} x^2 c^8-16 \r^6 x^2 c^6-2 \r^{12} x^2 c^4\\
&+\r^{12} x^2)\,,
\end{split}
\eqlabel{beomx2}
\end{equation}
where the prime now denotes a derivative with respect to $x$,
and we further introduced 
\begin{equation}
\r\equiv e^\a\,,\qquad c\equiv e^\chi\,.
\eqlabel{defrc}
\end{equation}

We demand that a physical RG flow should correspond to a background
geometry without naked singularities.  To ensure regularity, it is
necessary to impose the following asymptotic conditions:
\nxt  at the origin,
\ie as $x\to 0_+$,
\begin{equation}
\r=r^o_0+\calo(x^2)\,,\qquad c=c^o_0+\calo(x^2)\,,
\eqlabel{ir}
\end{equation}
with $r^o_0 c^o_0 \ne 0$;
\nxt and at the boundary, \ie as $y\equiv \frac 1x\to 0_+$,
\begin{equation}
\begin{split}
&\r=1+ y^2\ \left(r^b_{1,0}+r^b_{11} \ln y\right)
+\calo(y^4\ln^2 y)\,,\\ 
&c=1+y\ c^b_{1,0}+\frac 12\ y^2\ (c^b_{1,0})^2+
y^3\ c^b_{1,0}\ \left(c^b_{2,0}+\left(4+\frac43(c^b_{1,0})^2\right)\ \ln y\right)
+\calo(y^4\ln y)\,.
\end{split}
\eqlabel{uv}
\end{equation}
The asymptotic coefficients $c_{1,0}^b$ and $r^b_{1,1}$ 
are related to masses of the 
bosonic and fermionic components of $\caln=2^*$ hypermultiplet.
The precise relation can be established as in
\cite{bdkl} (see Appendix \ref{susya} for specific details):
\begin{equation}
c_{1,0}^b=\frac k2\,,\qquad r^b_{1,1}=\frac 16\ k^2\,,\qquad k\equiv m L\,.
\eqlabel{pwidt}
\end{equation} 
Given $k$ (and thus $c_{1,0}^b$ and $r^b_{1,1}$  via \eqref{pwidt} ) there is 
a unique nonsingular RG flow specified by 
\begin{equation}
\{r_0^o\,, c_0^o\,, r^b_{1,0}\,, c_{2,0}^b\}\,.
\eqlabel{parms}
\end{equation}

\begin{figure}[t]
\begin{center}
\psfrag{k}{{$k$}}
\psfrag{r0}{{$r_0^o$}}
\psfrag{c0}{{$\left(c_0^o-1-\frac 16 k \right)$}}
  \includegraphics[width=3in]{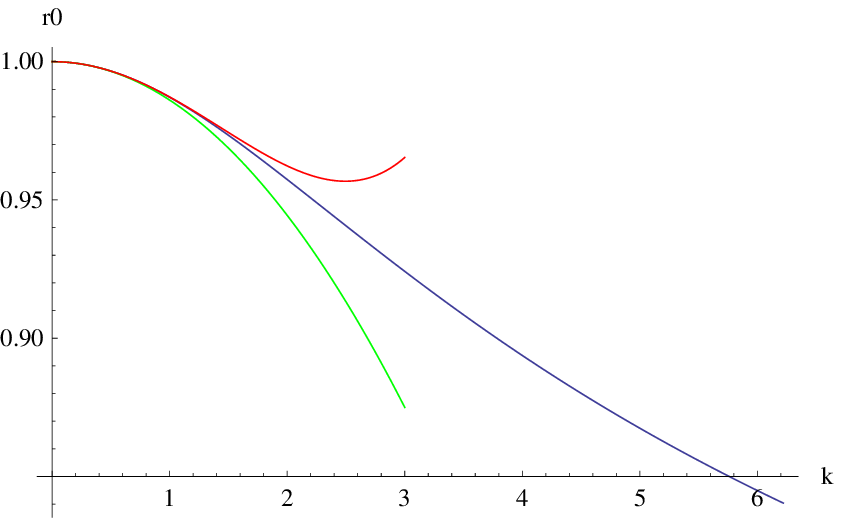}
  \includegraphics[width=3in]{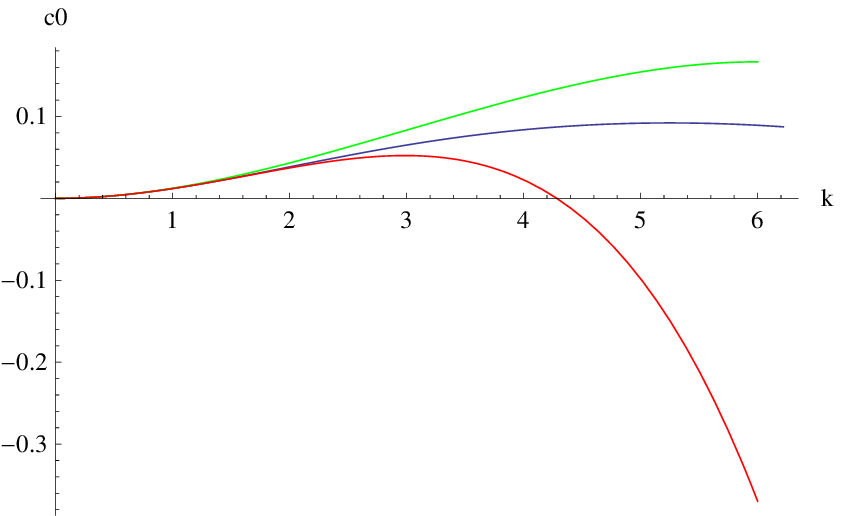}
\end{center}
  \caption{(Color online) RG flow coefficients $\{r_0^o, c_0^o\}$ 
as a function of $k$ (blue curves). The green curves represent perturbative 
approximations \eqref{pertsolpar} to orders $\calo(k^2)$ and $\calo(k^3)$
correspondingly; the red curves represent full perturbative 
approximations \eqref{pertsolpar}.   
}\label{figure1}
\end{figure}

\begin{figure}[t]
\begin{center}
\psfrag{k}{{$k$}}
\psfrag{r10}{{$r_{1,0}^b$}}
\psfrag{c20}{{$\left(c_{2,0}^b-2 \right)$}}
  \includegraphics[width=3in]{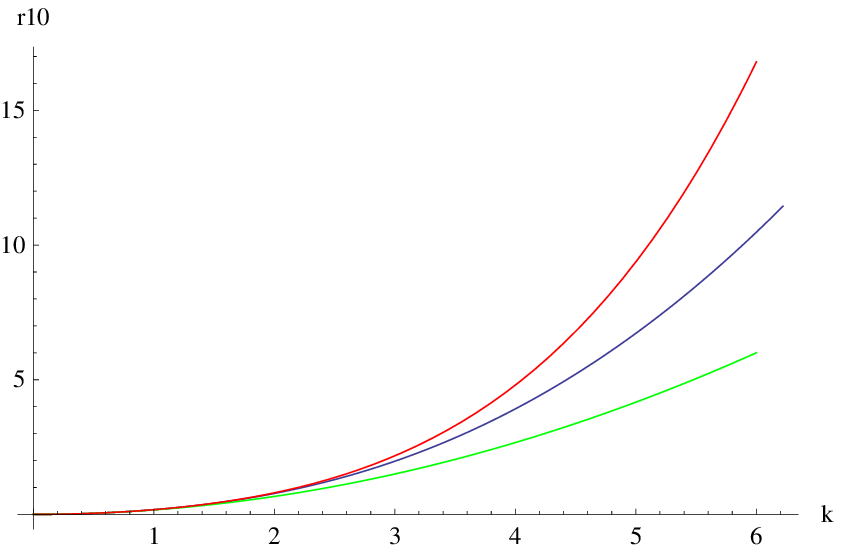}
  \includegraphics[width=3in]{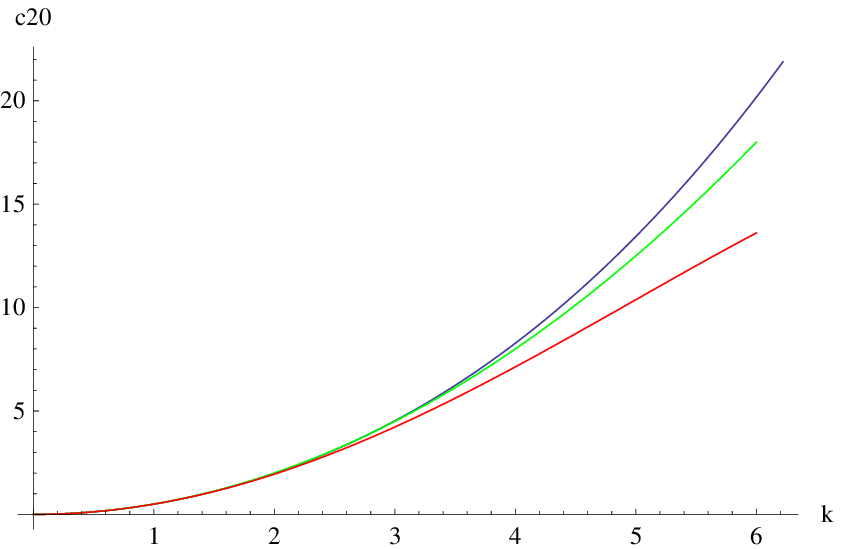}
\end{center}
  \caption{(Color online) RG flow coefficients $\{r_{1,0}^b, c_{2,0}^b\}$ 
as a function of $k$ (blue curves). The green curves represent perturbative 
approximations \eqref{pertsolpar} to order $\calo(k^2)$; 
the red curves represent full perturbative 
approximations \eqref{pertsolpar}.   
}\label{figure2}
\end{figure}

While it is difficult to construct analytic solutions to \eqref{beomx1}-\eqref{uv}
for generic $k$, and thus determine \eqref{parms}, 
it is possible to do so perturbatively in $k$. We 
find\footnote{We illustrate the solution to order $\calo(k^2)$ inclusive in 
Appendix \ref{pertsolution}.}
\begin{equation}
\begin{split}
&r_0^o=1-\frac{1}{72} k^2+\frac{1}{16} k^4\ r^o_{0,4}+\calo(k^6)\,,\qquad 
r_{0,4}^o=0.0178(5)\,,\\
&c_0^o=1+\frac16 k+\frac{1}{72} k^2-\frac{1}{648} k^3
-\frac{11}{31104} k^4+\frac{1}{32} k^5\ c^o_{0,5}+\calo(k^6)\,,
\qquad c_{0,5}^o=-0.0003(2)\,,\\
&r_{1,0}^b=\frac16 k^2+\frac{1}{16} k^4\ r_{1,0,4}^b+\calo(k^6)\,,
\qquad r_{1,0,4}^b=0.1333(3)\,,\\
&c_{2,0}^b=2+\frac12 k^2+\frac{1}{16} k^4\ c_{2,0,5}^b+\calo(k^6)\,,
\qquad c_{2,0,5}^b=-0.0542(2)\,.
\end{split}
\eqlabel{pertsolpar}
\end{equation}
Furthermore, the solution can always be found numerically, using the 'shooting method' 
introduced in \cite{abk}. Results of the latter numerical analysis are presented in
Fig.~\ref{figure1} and Fig.~\ref{figure2}.

\subsection{$\calf$ via holographic renormalization}
Following gauge-gravity correspondence, the free energy $\calf$ of the boundary 
theory is given by the Euclidean gravitational action of its holographic dual $S_E$. 
As usual, ultraviolet divergences in the field theory in computing $\calf$ 
are reflected in the infrared divergences of the dual gravitational bulk geometry.
Both must be regularized and renormalized. In the context of PW flow
the holographic renormalization has discussed in details in \cite{n2hydro,Buchel:2012gw}.
Below, we present the necessary details.

Let $r_c$ be the position of the boundary, and $S_E^{r_c}$ 
be the Euclidean gravitational action 
on the cut-off space 
\begin{equation}
\lim_{r_c\to \infty} S_E^{r_c}= S_E\,,
\eqlabel{cutac}
\end{equation}
where  $S_E$ is the on-shell Euclidean version of \eqref{action5}.
Using equations of motion \eqref{beom},
the regularized action takes form
\begin{equation}
S_{E}^{r_c}=\frac{1}{4\pi G_5}\biggl(\frac 18 \left(c_1(r)^4\right)'
\bigg|_0^{r_c}-\frac 32 \int_0^{r_c}c_1(r)^2\ dr \biggr){\rm vol}(S^4)\,,
\eqlabel{regaction}
\end{equation}
Notice that the integral contribution in \eqref{regaction}
arises entirely from  $S^4$ curvature. 
Besides the standard Gibbons-Hawking term 
\begin{equation}
S_{GH} = -\frac{1}{8\pi G_5} {\rm vol}(S^4) \sqrt{h_E}
\nabla_{\mu} n^{\mu} = -\frac{1}{8\pi G_5} \bigg[\left(c_1(r)^4\right)'
\bigg]\bigg|^{r_c}{\rm vol}(S^4)\,,
\eqlabel{gh}
\end{equation}
we supplement the combined 
regularized action $\left(S_E^{r_c}+S_{GH}\right)$ by the appropriate boundary 
counterterms which are needed to get a finite action. These boundary 
counterterms must be constructed from the local
metric and $\{\a=\ln\r,\chi\}$ scalar invariants on the boundary $\del\calm_5$,
except for the terms associated with the conformal anomaly  which
include an explicit dependence on the position of the boundary 
\cite{n2hydro,Buchel:2012gw},
\begin{equation}
\begin{split}
S^{counter}&=\frac{1}{4\pi G_5}{\rm vol}(S^4)\ c_1(r_c)^4\biggl[
\frac 34+\frac14\ R_{S^4}
+\frac12\ \chi^2+3\ \a^2+\frac32\ 
\frac{\a^2}{\ln \e}\\
&+ \ln \e\ \left(\frac 13\ R_{S^4}\ \chi^2 + \frac23\ \chi^4
-\frac 12 \left(R_{S^4\ ij}R_{S^4}^{ij}-\frac 13\ R_{S^4}^2\right)\right)
+\call_{ambiguity}
\biggr]\,,\\
\call_{ambiguity}&=
\dd_1\ \frac{\a^2}{\ln^2\e}+\dd_2\ \chi^4+\dd_3\ R_{S^4}\ \chi^2
+\dd_4\ R_{S^4}^2+\dd_5\ \frac{\a}{\ln\e}\ R_{S^4}  \,.
\end{split}
\eqlabel{scount1}
\end{equation}
Here $R_{S^4}$ and $R_{S^4\ ij}$ are the  $S^4$ Ricci scalar and tensor;
the coefficients $\{\dd_1\cdots \dd_5\}$ parameterize 
ambiguities of the holographic renormalization scheme. 
The conformal anomaly terms depend on the position of the boundary; we
choose to parameterize this position by the physical quantity
\begin{equation}
\e\equiv \frac{1}{\sqrt{g_{S^4S^4}}}\bigg|^{r_c}=c_1(r_c)^{-1}\,.
\eqlabel{eedef}
\end{equation} 
 The counterterms \eqref{scount1} are fixed in such a way 
that the {\it renormalized} Euclidean action $I_E$ is finite 
\begin{equation}
I_E\equiv \lim_{r_c\to \infty}\ \biggl(
S_E^{r_c}+S_{GH}+S^{counter}\biggr)\,,\qquad |I_E|<\infty\,.
\eqlabel{IEdef1}
\end{equation}

In what follows we construct each term in \eqref{IEdef1} explicitly. Since the 
RG flow is parameterized by $x$, \eqref{radgauge}, we use the same coordinate
in computing $I_E$. Starting from \eqref{backconst}, 
\begin{equation}
\begin{split}
0=&\biggl(\left(\frac{\r'(x)}{\r(x)}\right)^2
+\frac 13\ \left(\frac{c'(x)}{c(x)}\right)^2-1\biggr)\left(\frac{dx(r)}{dr}\right)^2-
\frac13\calp(\r(x),c(x))+\frac {1}{x^2}\,,
\end{split}
\eqlabel{dxdr1}
\end{equation}
where the prime denotes derivative with respect to $x$,
we find
\begin{equation}
\begin{split}
\left(\frac {dx(r)}{dr}\right)^2=&\left(3\r(x)^2 c(x)^2-3x^2c(x)^2 (\r'(x))^2-x^2\r(x)^2 (c'(x))^2\right)^{-1} \\
&\times \biggl(
\frac14 \rho(x)^4 x^2+3 \rho(x)^2 c(x)^2+\frac{c(x)^2 x^2}{4\rho(x)^2}
-\frac{c(x)^6 \rho(x)^{10} x^2}{64}
+\frac14 c(x)^4 \rho(x)^4 x^2\\
&+\frac{c(x)^2 \rho(x)^{10} x^2}{32}
-\frac{\rho(x)^{10} x^2}{64c(x)^2}
\biggr)\,.
\end{split}
\eqlabel{dxdr2}
\end{equation}
Denoting $x_c=x(r_c)$ and using $vol(S^4)=8\pi^2/3$, we rewrite \eqref{regaction} as 
\begin{equation}
\begin{split}
&S_E^{r_c}=\frac{2\pi}{3 G_5}\biggl(\calj_1+\calj_2\biggr)\,,\\
&\calj_1\equiv \frac 12\ x^3\ \left(\frac{dx(r)}{dr}\right)\bigg|_{0}^{x_c}\,,\qquad
\calj_2\equiv -\frac 32 \int_0^{x_c}\ x^2 \left(\frac{dx(r)}{dr}\right)^{-1}\ dx\,.
\end{split}
\eqlabel{bulkSE}
\end{equation}
Notice that using \eqref{dxdr2}, both $\calj_1$ and $\calj_2$ are expressed through $x$ radial 
coordinate. Also, given \eqref{eedef}, $\e=x_c^{-1}$.
Using asymptotic expansions \eqref{ir} and \eqref{uv} we find
\begin{equation}
\begin{split}
&\calj_1\equiv \calj^{singular}_1+\calj_1^{finite}\,,\\
&\calj^{singular}_1=\frac 14\ \e^{-4}+\left(\frac 12+\frac 16 (c_{1,0}^b)^2\right)\ \e^{-2}+(r_{1,1}^b)^2\ \ln^2\e+
\biggl(2 (c_{1,0}^b)^2+\frac23 (c_{1,0}^b)^4\\
&+2r_{1,1}^b r_{1,0}^b
+\frac12 \left(r_{1,1}^b\right)^2\biggr)\ \ln\e\,,\\
&\calj_1^{finite}=\frac12 r_{1,0}^b r_{1,1}^b+\frac12 (c_{1,0}^b)^2 c_{2,0}^b+\frac18
 (r_{1,1}^b)^2+(r_{1,0}^b)^2+\frac{1}{36} (c_{1,0}^b)^4+\frac16 (c_{1,0}^b)^2-\frac12\\
&+\calo(\e^2\ln^3\e)\,,
\end{split}
\eqlabel{j1}
\end{equation}
where we explicitly separated the singular and the finite parts of 
$\calj_1$ as $\e=x_c^{-1}\to 0$. Note that both $\calj^{singular}_1$ and $\calj^{finite}_1$ receive 
contribution only from the boundary. Unlike $\calj_1$, $\calj_2$ can not be computed in 
closed form analytically. Here we find: 
\begin{equation}
\begin{split}
&\calj_2\equiv \calj^{singular}_2+\calj_2^{finite}\,,\\
&\calj^{singular}_2=-\frac32 \int_1^{1/\e} dx 
\left(2x-\frac{4(c_{1,0}^b)^2+12}{3x}\right)=-\frac{3}{2}\e^{-2}
-(2(c_{1,0}^b)^2+6)\ln\e +\frac 32\,,\\
&\calj_2^{finite}=-\frac 32 \int_0^1 x^2 \left(\frac{dx(r)}{dr}\right)^{-1}\ dx
-\frac32 \int_1^{1/\e} dx \biggl(x^2 \left(\frac{dx(r)}{dr}\right)^{-1}
\\
&-\left(2x-\frac{4(c_{1,0}^b)^2+12}{3x}\right)\biggr)\,.
\end{split}
\eqlabel{j2}
\end{equation}
Next, we represent GH term \eqref{gh} as 
\begin{equation}
\begin{split}
S_{GH}=\frac{2\pi}{3G_5}\ \calj_3\,,\qquad \calj_3=-2 x^3 
\left(\frac{dx(r)}{dr}\right)\bigg|^{x_c}\,.
\end{split}
\eqlabel{j3}
\end{equation} 
Using \eqref{ir} and \eqref{uv} we find 
\begin{equation}
\begin{split}
&\calj_3\equiv\calj_3^{singular}+\calj_3^{finite}\,,\\
&\calj_3^{singular}=-\e^{-4}-\left(\frac 23 (c_{1,0}^b)^2+2\right)\e^{-2}
-4 (r_{1,1}^b)^2\ \ln^2\e\\
&-\left(8r_{1,0}^br_{1,1}^b+2(r_{1,1}^b)^2+8(c_{1,0}^b)^2
+\frac 83(c_{1,0}^b)^4\right)\ \ln\e\,,\\
&\calj_3^{finite}=-2 r_{1,0}^b r_{1,1}^b-2 (c_{1,0}^b)^2 c_{2,0}^b
-\frac23 (c_{1,0}^b)^2-\frac19 (c_{1,0}^b)^4-\frac12 (r_{1,1}^b)^2-4 (r_{1,0}^b)^2
+2\\
&+\calo(\e^2\ln^3\e)\,.
\end{split}
\eqlabel{j3sf}
\end{equation}
Finally, 
\begin{equation}
\begin{split}
&S^{counter}=\frac{2\pi}{3G_5}\ \calj_4\,,\qquad 
\calj_4\equiv\calj_4^{singular}+\calj_4^{finite}+\calj_4^{finite,ambiguity}\,,\\ 
&\calj_4^{singular}=\frac 34 \e^{-4}+\left(3+\frac 
12 (c_{1,0}^b)^2\right)\e^{-2}+3(r_{1,1}^b)^2\ \ln^2\e\\
&+\left(8(c_{1,0}^b)^2+2(c_{1,0}^b)^4+6+\frac32 (r_{1,1}^b)^2
+6r_{1,0}^br_{1,1}^b
\right)\ \ln\e\,,\\
&\calj_4^{finite}=3 (r_{1,0}^b)^2+(c_{1,0}^b)^2 c_{2,0}^b
-\frac16 (c_{1,0}^b)^4+3 r_{1,0}^b r_{1,1}^b+\calo(\ln^{-1}\e)\,,\\
&\calj_4^{finite,ambiguity}=\dd_1 (r_{1,1}^b)^2+\dd_2 (c_{1,0})^4
+12 \dd_3 (c_{1,0})^2+144 \dd_4+\calo(\ln^{-1}\e)\,.\\
\end{split}
\eqlabel{j4}
\end{equation}
Note that irrespectively as to whether or not supersymmetry is preserved 
along the RG flow, \ie $r_{1,1}^b=\frac 23 (c_{1,0}^b)^2$ --- 
see \eqref{pwidt}, all the singularities in $I_E$ cancel:
\begin{equation}
\calj_1^{singular}+\calj_2^{singular}+\calj_3^{singular}+\calj_4^{singular}=\frac32\,.
\eqlabel{finalsing}
\end{equation}
Collecting the finite pieces (and accounting for \eqref{finalsing}),  
we find
\begin{equation}
\begin{split}
I_E=&\frac{2\pi}{3 G_5}\biggl(3+\frac32 r_{1,0}^b r_{1,1}^b
+\left(\dd_1-\frac 38\right) 
(r_{1,1}^b)^2+(c_{1,0}^b)^2 \left(12\dd_3-\frac12 c_{2,0}^b-\frac12\right)
\\
&+\left(\dd_2-\frac14\right) (c_{1,0}^b)^4+144 \dd_4-12 \dd_5\ r_{1,1}^b+\lim_{\e\to 0} 
\calj_2^{finite}\biggr)\,,
\end{split}
\eqlabel{finalIE}
\end{equation}
where $\dd_i$ are renormalization scheme-dependent ambiguities.
For supersymmetric RG flows (see \eqref{pwidt}) we have
\begin{equation}
\begin{split}
I_E^{susy}=&\frac{2\pi}{3 G_5}\biggl(
3+\left(-\frac 18-\frac18 c_{2,0}^b
+\frac14 r_{1,0}^b\right) k^2-\frac{5}{192}k^4+\lim_{\e\to 0}
\calj_2^{finite}\biggr)\\
&+I_{E}^{susy,ambiguity}\,,\qquad 
I_{E}^{susy,ambiguity}=
\frac{2\pi}{3 G_5}(\cala_0+\cala_2 k^2+\cala_4 k^4) \,,
\end{split}
\eqlabel{finalIEsusy}
\end{equation}
where $\cala_i$ are ambiguity (purely numerical --- mass 
independent) coefficients. Notice that an immediate consequence of 
\eqref{finalIEsusy} is that from the dual gravitational perspective,
$I_E^{susy}$ is ambiguous up to an order-two polynomial in $(m L)^2$. 

In section \ref{solution} we constructed perturbative in $k$, and fully 
nonlinear in $k$, gravitational RG flows corresponding to compactification of 
$\caln=2^*$ gauge theory on $S^4$. In the rest of this section we outline perturbative 
computations of $I_{E}^{susy}$ to order $\calo(k^2)$, present perturbative results to 
order $\calo(k^6)$ inclusive\footnote{As we emphasized in \eqref{finalIEsusy} 
this is the first renormalization scheme-independent 
contribution to $I_E^{susy}$.}, and also present full nonperturbative in $k$ results.

Using \eqref{c1res}-\eqref{r2res} and \eqref{dxdr2}, we find
\begin{equation}
\begin{split}
&-\frac 32 \int_0^1 x^2 \left(\frac{dx(r)}{dr}\right)^{-1}=
-\frac{3\sqrt{5}}{2}+6\ \arcsinh\frac12
+\left(\frac k2\right)^2\biggl(
-\frac{24}{\sqrt{5}}\ \arctanh^2\frac{1}{\sqrt{5}}
\\
&+12\ \arctanh\frac{1}{\sqrt{5}}
-\frac{19\sqrt{5}}{10}+2\ \arcsinh\frac12
\biggr)+\calo(k^4)\,,\\
&\lim_{\e\to 0}\biggl[-\frac32 \int_1^{1/\e} dx \left(x^2 
\left(\frac{dx(r)}{dr}\right)^{-1}
-\left(2x-\frac{4(c_{1,0}^b)^2+12}{3x}\right)\right)\biggr]=
-\frac 92 +\frac{3\sqrt{5}}{2}\\
&-6\ \arctanh\frac{1}{\sqrt{5}}
+\left(\frac k2\right)^2\biggl(
\frac{24}{\sqrt{5}}\ \arctanh^2\frac{1}{\sqrt{5}}-14\ 
\arctanh\frac{1}{\sqrt{5}}-\frac 72+\frac{19\sqrt{5}}{10}
\biggr)+\calo(k^4)\,,
\end{split}
\end{equation}
leading to 
\begin{equation}
\lim_{\e\to 0}\calj_2^{finite}=-\frac 92 -\frac72\ \left(\frac k2\right)^2\,.
\eqlabel{j2tot}
\end{equation}
Further using \eqref{finalIEsusy} we find
\begin{equation}
\begin{split}
I_E^{susy}=&\frac{2\pi}{3 G_5}\biggl(
-\frac 32 -5\ \left(\frac k2\right)^2 +\calo(k^4) \biggr)
+I_{E}^{susy,ambiguity}\,.
\end{split}
\eqlabel{finalIEsusyper}
\end{equation}
A straightforward, albeit tedious computation extends \eqref{finalIEsusyper}
to order $\calo(k^6)$:
\begin{equation}
\begin{split}
I_E^{susy}=&\frac{2\pi}{3 G_5}\biggl(
-\frac 32 -5\ \left(\frac k2\right)^2-\frac34\ \left(\frac k2\right)^4
+\cali_6\  \left(\frac k2\right)^6 +\calo(k^8) \biggr)
+I_{E}^{susy,ambiguity}\,,\\
\cali_6=&0.10696(3)\,.
\end{split}
\eqlabel{finalIEsusyper1}
\end{equation}
From \eqref{finalIEsusyper1}, the leading unambiguous contribution 
to $I_E^{susy}=\calf$ 
is\footnote{We use \eqref{g5}, and $L=2$, $R=1$ to express the 
results in gauge theory variables.} 
\begin{equation}
\calf=\frac{N^2}{6}\biggl({\rm ambiguous}+\cali_6 (m R)^6
+\calo((m R)^8)\biggr)\,.
\eqlabel{pertffin}
\end{equation}

\begin{figure}[t]
\begin{center}
\psfrag{mr}{{$(m R)^2$}}
\psfrag{f}{{$\frac{1}{N^2}\ \calf $}}
\psfrag{c0}{{$\left(c_0^o-1-\frac 16 k \right)$}}
  \includegraphics[width=3in]{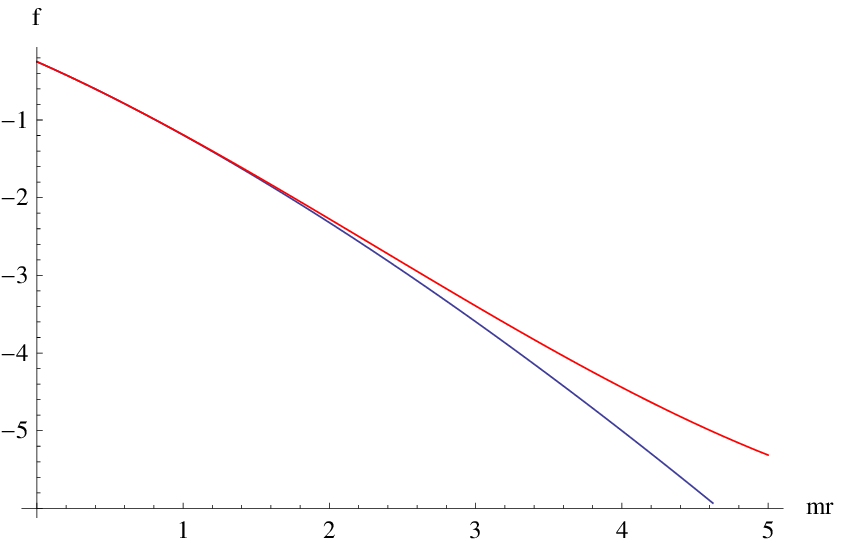}
  \includegraphics[width=3in]{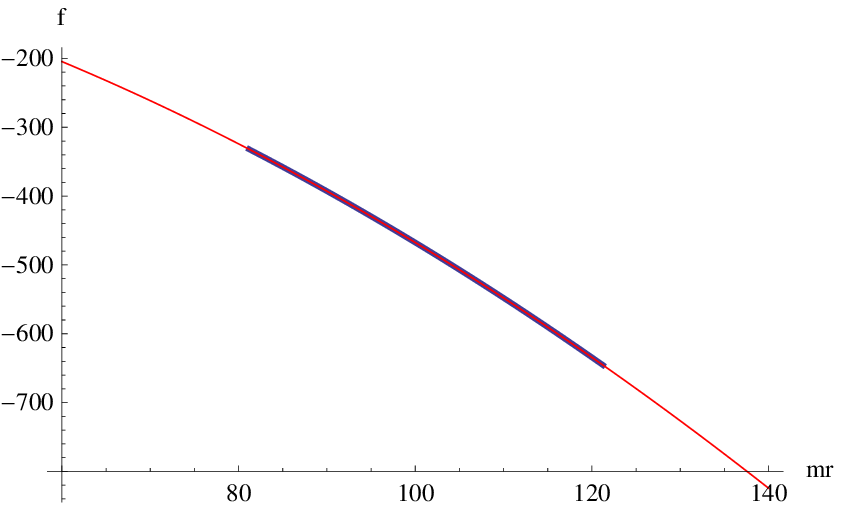}
\end{center}
  \caption{(Color online) Blue curves represent 
free energy $\calf$ of the holographic PW flow compactified on 
$S^4$ of radius $R$, see \eqref{finalIEsusy}, in the scheme with 
$\cali_{E}^{susy,ambiguity}=0$.  The red curve in the {\it left panel} is 
perturbative in $(m R)^2$ approximation to $\calf$, 
see \eqref{finalIEsusyper1}. The red curve in the {\it right panel} 
is the best fit to data with the ansatz \eqref{ffit2}.   
}\label{figure3}
\end{figure}

Given numerical solution to RG flow of $\caln=2^*$
gauge theory on $S^4$ as discussed in section \ref{solution},
we can evaluate its free energy \eqref{finalIEsusy} in the scheme with 
$\cali_{E}^{susy,ambiguity}=0$. Results are presented in Fig.~\ref{figure3}.
Blue curves represent $\calf $ as a function of $(m R)^2$.
The red curve in the {\it left panel} is 
perturbative in $(m R)^2$ approximation to $\calf$, 
see \eqref{finalIEsusyper1}. A polynomial fit of 30 first
points to a blue curve produces 
\begin{equation}
\frac {6}{N^2}\ \calf \bigg|_{fit}=-1.5-5.0 (m R)^2 -0.749996 (mR)^4
+0.106829 (m R)^6+\calo((m R)^8) \,,
\eqlabel{ffit}
\end{equation}
in excellent agreement with  \eqref{finalIEsusyper1}.
The red curve in the {\it right panel} is the matrix model 
motivated fit to data (see \eqref{fn2eq4}):
\begin{equation}
\begin{split}
&\frac {1}{N^2}\ \calf \bigg|_{fit}=\calf_0+\calf_1\ (mR)^2+\calf_2\ (mR)^4+
\calf_3\ (1+(mR)^2)\ln \left(1+(mR)^2\right)\,,\\
&\calf_0=-2.00(7)\,,\qquad \calf_1=-0.18(3)\,,\qquad \calf_2=-0.02(8)\,,\qquad \calf_3=-0.36(9)\,.
\end{split}
\eqlabel{ffit2}
\end{equation}
We used 2000 points in the fit.
 
We would like to compare \eqref{ffit} and \eqref{ffit2} with the matrix model result \eqref{fn2eq4}.  
As we emphasized earlier, while the holographic free energy $\calf$ is ambiguous, these ambiguities
are completely parameterized by a second-order polynomial in $(m R)^2$. A choice of the latter 
polynomial is equivalent to a choice of the renormalization scheme. Thus, for $mR \ll 1$
the leading unambiguous coefficient is that of $(mR)^6$, \ie $N^2 \cali_6/6$. Since 
\begin{equation}
\frac{1}{6}\ \cali_6\ \ne \ \frac{1}{12}\,,
\eqlabel{smallr}
\end{equation}   
where the RHS is the matrix model prediction \eqref{fneq5},  we conclude that 
$S^4$-compactified PW flow can not represent a holographic dual to 
Pestun's large-$N$ matrix model \cite{Pestun:2007rz} for arbitrary 
values of $mR$. 

Is it possible to recover the holographic result in the $S^4$ decompactification limit?
Holographic renormalization scheme generically differs  from the 
scheme implicit in the matrix model computation. To account for possible 
differences, we fit the large $mR$ data points of the holographic 
free energy with the ansatz \eqref{ffit2}. Notice that the coefficients 
$\{\calf_0,\cdots \calf_2\}$ encode the generic scheme dependence, on top of the expected result 
$\calf_{\caln=2^*}^{loc}$, \eqref{fn2eq4}. Since 
\begin{equation}
\calf_3\ne -\frac 12\,,
\eqlabel{larger}
\end{equation}  
where the RHS is the $mR\gg 1$ matrix model prediction \eqref{fneq5}, 
we conclude that Pestun's large-$N$ matrix model \cite{Pestun:2007rz} 
can not reproduce the decompactification limit of the PW flow on $S^4$.

\section{Conclusion}\label{section4}
In this paper we address the question whether matrix model localization method developed by Pestun 
for $\caln=2$ gauge theories can be used to explain the selection of the Coulomb branch vacuum 
in holographic $\caln=2$ renormalization group flows. We focus on two examples:
the $\caln=2^*$ RG flow \cite{pw}, and the RG flow to the SYM \cite{Gauntlett:2001ps}.
While in the former case the matrix model correctly identifies the PW vacuum \cite{Buchel:2013id},
the matrix model analysis \cite{Russo:2012ay} fails to reproduce the GKMW vacuum --- 
there is an agreement though in the extreme high-energy limit, \ie at energy scales much higher 
than the compactification scale set by the $S^4$ radius. 
A possible reason for the discrepancy in the SYM case can be attributed to the 
fact that the matrix model saddle-point in the $S^4$ decompactification limit 
is unphysical, as it is identified in the regime with negative running $g_{YM}^2$ coupling.  

Further detailed comparison demonstrates that the free energy of the $S^4$ compactified 
PW flow presented here disagrees with the matrix model computation reported in \cite{Buchel:2013id}.
The disagreement occurs both for $mR\ll 1$ and in the decompactification limit $m R\gg 1$.
We argued that disagreement can not be an artifact of the (potential) difference in 
holographic and matrix model renormalization schemes.  

It is important to understand a reason why the five-dimensional PW effective action \cite{pw}, 
which represents a consistent truncation of type IIb supergravity \cite{b1}, and agrees (at least in the 
asymptotic boundary region)
with the holographic rules of constructing massive deformations of $\caln=4$ SYM, apparently, 
does not contain a holographic flow dual to Pestun's $\caln=2^*$ $S^4$ supersymmetric compactification.
If there is a different (unknown as advocated in \cite{Russo:2013qaa}) gravitational dual, 
how precisely is it different from the PW model? We believe that resolving this issue will
lead to a deeper understanding of non-conformal holographic RG flows.

~\\
\section*{Acknowledgments}
It is a pleasure to thank Jaume Gomis, Jorge Russo and Kostya Zarembo
for valuable discussions. This work was supported by NSERC through Discovery Grant. 
Research at Perimeter
Institute is supported through Industry Canada and by the Province of Ontario
through the Ministry of Research \& Innovation.

\appendix

\section{Matching $\caln=2^*$ masses to SUGRA RG flow non-normalizable 
coefficients}\label{susya}

PW solution \cite{pw} represents a gravitational dual to 
$\caln=2^*$ gauge theory on $R^{3,1}$. Given appropriate  metric ansatz,
\begin{equation}
ds_5^2=c_1(r)^2 (dR^{3,1})^2+dr^2\,,
\eqlabel{susy}
\end{equation} 
solutions of supersymmetric RG flows from 
effective action \eqref{action5}-\eqref{supp} can be parameterized as 
\cite{pw}
\begin{equation}
\begin{split}
c_1&=\frac{k \r^2}{\sinh(2\chi)}\,,\\
\r^6&=\cosh(2\chi)+\sinh^2(2\chi)\,\ln\frac{\sinh(\chi)}{\cosh(\chi)}\,,
\end{split}
\eqlabel{pwsolution}
\end{equation}
where the single integration constant $k$ is related to the hypermultiplet
mass $m$ according to \cite{bpp}
\begin{equation}
k= m L =2 m\,.
\eqlabel{kim}
\end{equation}
Introducing a new radial coordinate as in \eqref{radgauge},
the leading boundary, \ie $y\equiv \frac 1x\to 0_+$,  
asymptotics of the flows \eqref{pwsolution} are given 
by  
\begin{equation}
\begin{split}
\r=&1+\frac{1}{12}\ k^2\ \left (2\ln\frac k2+1+ 2\ln y \right )y^2+\calo(y^4 \ln^2 y)\,,\\
e^\chi\equiv c=&1+\frac 12\ y\ k+\frac18\ k^2\ y^2+y^3\ k^3\ 
\frac{1}{48}\left(8\ln\frac k2+1+8\ln y\right)+\calo(y^4\ln y)\,.
\end{split}
\eqlabel{uvpw}
\end{equation}
Matching \eqref{uvpw} with \eqref{uv} we identify
\begin{equation}
c_{1,0}^b=\frac k2\,,\qquad r^b_{1,1}=\frac 16\ k^2\,.
\eqlabel{pwid}
\end{equation} 

\section{Perturbative solution of \eqref{beomx1}-\eqref{uv} to order $\calo(k^2)$}
\label{pertsolution}

It is straightforward to solve \eqref{beomx1}-\eqref{uv} perturbatively 
in $\calo(k^2)$.  Specifically, assuming 
\begin{equation}
c(y)=1+\sum_{n=1}^\infty\ \left(\frac{k}{2}\right)^n\ c_{(n)}(y)\,,\qquad 
\r(y)=1+\sum_{n=1}^\infty\ \left(\frac{k^2}{6}\right)^n\ \r_{(2n)}(y) \,,
\end{equation}
with the asymptotic boundary conditions 
\begin{equation}
c_{(n)}(y)=\dd^1_n\ y+\calo(y^2)\,,\qquad \r_{(2n)}(y) =\dd^1_n\ y^2 \ln y+\calo(y^2)\,,
\eqlabel{pertuv}
\end{equation}
and regularity as $x\to 0_+$, we find from \eqref{beomx1}-\eqref{beomx2}
two sets of ODEs:
\begin{equation}
\begin{split}
0=&c_{(n)}''-\frac{3+8y^2}{y(1+4y^2)}\ c_{(n)}'+\frac{3}{y^2(1+4y^2)}\ c_{(n)}+\cals_{c,n}\,,\\
0=&\r_{(2n)}''-\frac{3+8y^2}{y(1+4y^2)}\ \r_{(2n)}'+\frac{3}{y^2(1+4y^2)}\ \r_{(2n)}+\cals_{\r,n}\,,
\end{split}
\eqlabel{perturbativeeoms}
\end{equation} 
where the source terms at order $n$, \ie  $\cals_{c,n}$ and $\cals_{\r,n}$, are
functionals of solutions at previous orders. To order $\calo(k^2)$ inclusive 
we find 
\begin{equation}
c_{(1)}=y\ \sqrt{1+4y^2}-4\ y^3\ \arctanh \frac{1}{\sqrt{1+4y^2}}\,,
\eqlabel{c1res}
\end{equation}
\begin{equation}
c_{(2)}=8\ y^6\ \arctanh^2 \frac{1}{\sqrt{1+4y^2}}-4\ y^4\sqrt{1+4y^2}\ 
\arctanh \frac{1}{\sqrt{1+4y^2}}+\frac 12 y^2+2y^4\,,
\eqlabel{c2res}
\end{equation}
\begin{equation}
\r_{(2)}=y^2- y^2\sqrt{1+4y^2}\ 
\arctanh \frac{1}{\sqrt{1+4y^2}}\,.
\eqlabel{r2res}
\end{equation}

\end{document}